\documentclass[9pt, twocolumn,nofootinbib]{revtex4-1}

\usepackage{graphicx} 
\usepackage[utf8]{inputenc}

\usepackage{color}

\usepackage{fontenc}

\usepackage{amsmath}
\usepackage{amssymb}
\usepackage{amsthm}
\usepackage{hyperref}
\usepackage{amsmath,amssymb,amsthm, graphicx, array,epsfig, exscale, dsfont, amsfonts, verbatim, fancyhdr, natbib, bbm,mathrsfs,textcomp,listings, epsfig}

\usepackage{color}

\usepackage{slashbox}

\newcommand{\one}{\mbox{$1 \hspace{-1.0mm}  {\bf l}$}}

\newcommand{\bra}[1]{\left\langle{#1}\right\vert}
\newcommand{\ket}[1]{\left\vert{#1}\right\rangle}

\newcommand{\etaD}{\eta_{\text{D}}}

\newcommand{\etaM}{\eta_{M}}

\date{\today}

\begin{document}

\title[Measurement-device-independent quantum key distribution with quantum memories]{Measurement-device-independent quantum key distribution with quantum memories}
\author{Silvestre Abruzzo, Hermann Kampermann, Dagmar Bru{\ss}}
\address{Institute for Theoretical Physics III, Heinrich-Heine-Universit\"at D\"usseldorf, Universitätsstr. 1, 40225 D\"usseldorf, Germany}


\begin{abstract}
We generalize measurement-device-independent quantum key distribution [ H.-K. Lo, M. Curty, and B. Qi, Phys. Rev. Lett. 108, 130503 (2012) ] to the scenario where the Bell-state measurement station contains also heralded quantum memories. We find analytical formulas, in terms of device imperfections, for all quantities entering in the secret key rates, i.e., the quantum bit error rate and the repeater rate. We assume either single-photon sources or weak coherent pulse sources plus decoy states. We show that it is possible to significantly outperform  the original proposal, even in presence of decoherence of the quantum memory. Our protocol may represent the first natural step for implementing a two-segment quantum repeater.
\end{abstract}

\maketitle

\newcommand{\mdirel}{MDI-QKD-RELAY }
\newcommand{\mdirels}{MDI-QKD-RELAY-SPS }
\newcommand{\mdirelw}{MDI-QKD-RELAY-WCP }

\newcommand{\mdirep}{MDI-QKD-REPEATER }
\newcommand{\mdireps}{MDI-QKD-REPEATER-SPS }
\newcommand{\mdirepw}{MDI-QKD-REPEATER-WCP }

\section{Introduction}
Quantum communication has been developed in the last thirty years. One prominent communication protocol is quantum key distribution (QKD) which aims at distributing a secret key between two distant parties. Suitable quantum systems for quantum communication are photons as they have very low decoherence and they can be easily generated, distributed and detected with standard technology. However, due to absorption in optical fibers (or free-space), QKD with reasonable rates is only possible up to ca. 150 km \cite{Scarani:2009}. To overcome this problem quantum repeaters have been developed \cite{briegel_quantum_1998}. The idea is to divide the distance between Alice and Bob in segments, to create entanglement in each segment and then to enlarge the distance using entanglement swapping. Nowadays, the constituting parts of a quantum repeater have been realized and small networks have been implemented  in a laboratory set-up \cite{sangouard_quantum_2011}. However, a complete quantum repeater (even with two 
segments) that will permit to outperform direct transmission has not been realized yet \cite{sangouard2012quantum}. 

Recently, measurement-device-independent QKD (\mdirel) has been proposed \cite{PhysRevLett.108.130503, PhysRevLett.108.130502}. This protocol is based on the principle of a quantum relay \cite{PhysRevLett.92.047904} and  uses weak coherent pulse (WCP) sources. Briefly speaking, two parties, Alice and Bob, each equipped with a WCP source, send photon pulses to a station which performs a Bell-state measurement (BSM) and communicates the result to Alice and Bob. Then Alice sends Bob information regarding the used basis  such  that if necessary Bob can implement a bit flip. This protocol is measurement-device-independent because Alice and Bob do not need to measure anything and therefore the protocol is immune to detector attacks \cite{lydersen2010hacking,gerhardt2011full}.   The \mdirel has already been  implemented experimentally both in laboratory environment and in a real-world environment \cite{2012arXiv1204.0738R,2012arXiv1207.6345F, 2012arXiv1209.6178L}. Moreover, more efficient protocols have already
been proposed \cite{PhysRevA.86.062319,PhysRevA.85.042307,2013arXiv1305.6965X} and finite-size corrections have been analyzed \cite{PhysRevA.86.022332, PhysRevA.86.052305,2013arXiv1305.6965X}.

In this paper we extend the original \mdirel protocol \cite{PhysRevLett.108.130503} introducing quantum memories in the BSM station. The first consequence is that heralding, provided by quantum memories, permits to improve the rate at a given distance where MDI-QKD can be used. The advantage of our protocol over other quantum repeater protocols is that it does not need entanglement sources but only commercial off-the-shelf weak coherent pulse sources. Quantum memories have not reached the commercial market yet but they are under active development. With our protocol we show that it is possible to use quantum memories with low coherence time.

The manuscript is organized as follows. In Sec.~\ref{sec:sps} we present a generalization of  measurement-device-independent QKD with single-photon sources to the scenario with quantum memories. We derive the formula for the secret key rate and we study its dependency on the decoherence of the quantum memories. Finally, we compare the secret key rate obtained with our protocol with the one obtained with the quantum relay proposed in \cite{PhysRevLett.108.130503}. In Sec.~\ref{sec:wcp} we generalize the whole analysis to WCP sources. In order to calculate the secret key rate we consider QKD with decoy states \cite{PhysRevLett.94.230504,PhysRevA.72.012326}. In Sec.~\ref{sec:conc} we give our conclusions.

\section{Scheme with single-photon states\label{sec:sps}}
In this section we extend the \mdirel protocol presented in \cite{PhysRevLett.108.130503} introducing quantum memories (QM) and using single-photon-sources (SPS), which would be the ideal type of source for this protocol. Therefore, although SPSs are still not practical they will permit to establish upper bounds on the achievable secret key rate, i.e.  sources with many-photon pulses or with additional imperfections will lead to a worse secret key rates. We denote the protocol considered in this section as \mdireps.

\begin{figure}
 \includegraphics[width=.45\textwidth]{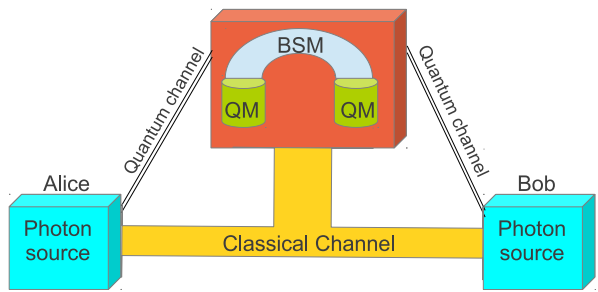}
\caption{(Color online) Scheme of a measurement-device-independent quantum repeater. The difference w.r.t. \mdirel is that quantum memories are used. QM=quantum memories, BSM=Bell-state measurement. The two sources produce single-photon states or weak coherent pulses.}
 \label{fig:scheme} 
\end{figure}

\subsection{The protocol}
In the following we give the steps of the protocol which is a generalization of the one proposed in \cite {PhysRevLett.108.130503} (see Fig.~\ref{fig:scheme}):
\begin{enumerate}
\item Alice and Bob prepare randomly and independently one of the four qubit states $\ket{\psi}\in\{\ket{0}, \ket{1}, \ket{+}, \ket{-}\}$ where $\ket{\pm}:=(\ket{0}\pm\ket{1})/\sqrt{2}$. We will refer to the set $\{\ket{0}, \ket{1}\}$ as the Z-basis (or rectilinear basis) and the set $\{\ket{+}, \ket{-}\}$ as the X-basis (or diagonal basis). The states are sent through the quantum channel to the repeater station. The information related to the created states is stored by Alice and Bob locally. This process is repeated continuously by Alice and Bob with frequency $\nu_s$ which is the repetition frequency of the source.  
\item When both quantum memories are filled up, the quantum memories are read and a Bell-state measurement (BSM) is performed. The result of the BSM and the fact that the measurement was successful are sent to both Alice and Bob.
\item If the measurement was successful Alice and Bob will keep their stored information and if needed one of the two parties will perform a bit flip. If the measurement was not successful then Alice and Bob will remove their classical information from their stored pool of data. 
\item After creating sufficiently many bits Alice and Bob do the usual QKD post-processing which consists of sifting, parameter estimation, error correction and privacy amplification \cite{Scarani:2009}.
\end{enumerate}

The second step is different from the original \mdirel protocol. Here quantum memories are used for increasing the entanglement swapping success probability. As a result the total secret key rate will be higher than for the case without quantum memories. 

\subsection{The secret key rate \label{sec:scrsps}}
Concerning the security, the protocol is equivalent to the entanglement-based repeater protocol \footnote{The equivalence is seen by the following arguments:  consider an entanglement-based repeater protocol where Alice and Bob each produce the state $\ket{\phi^+}_{AC}=\ket{\phi^+}_{DB}:=\frac{1}{\sqrt{2}}\left(\ket{00}+\ket{11}\right)$. The subsystems $C$ and $D$ are sent to the channel and subjected to a BSM. On the other hand, subsystems $A$ and $B$ remain in Alice's and Bob's laboratory and are measured in basis X or Z. For the case where both Alice and Bob have chosen basis $Z$, the measurement is described by two projectors $\{\Pi^{(0)}:=\ket{0}\bra{0},\Pi^{(1)}:=\ket{1}\bra{1}\}$. The resulting state is given by $\left((\Pi^{i}_{A}\otimes\Pi^{j}_{B})\otimes \mathcal{E}_{CD}\right) (\ket{\phi^+}_{AC}\otimes\ket{\phi^+}_{DB})$ with $i,j=0,1$. The QKD measurement and BSM act on different Hilbert spaces and therefore they can be interchanged leading to $\left(\mathcal{E}_{CD} \otimes (\Pi^{i}_{A}\otimes\
Pi^{j}_{B})\right) (\ket{\phi^+}_{AC}\otimes\ket{\phi^+}_{DB})=\mathcal{E}_{CD}(\ket{i}_{C}\otimes\ket{j}_{D})$ where the state $\ket{i}_{C}\otimes\ket{j}_{D}$ represents two single photons prepared in the  Z basis with polarization $i$ and $j$. The case of the $X$ basis is analogous.} \cite{bennett1984quantum, bennett_quantum_1992, PhysRevLett.108.130503}.  In this paper we consider the asymptotic secret key rate which gives an upper bound on the achievable secret key rate. Finite size corrections can be included using the analysis done in \cite{PhysRevA.86.022332, PhysRevA.86.052305}. The formula for the asymptotic secret key rate is given in \cite{Scarani:2009, PhysRevLett.108.130503}
\begin{equation}
\label{eq:kr:sps}
r_{\infty}^{\textrm{REP}}:=\frac{1}{<T>}(1-h(e_Z)-h(e_X)),
\end{equation}
where $h(p):=-p\log_2p-(1-p)\log_2(1-p)$ is the binary Shannon entropy, $e_X$($e_Z$) is the quantum bit error rate (QBER) in the $X$-basis ($Z$-basis) and $\frac{1}{<T>}$ is the raw key rate\footnote{The sifting rate does not appear because we employ an asymmetric  protocol where Alice and Bob produce with probability almost one a state in base X and the remaining times a state in base Z \cite{Lo2005}.}. The QBER represents the fraction of discordant bits in the raw key, which is the collection of bits stored by Alice and Bob before the post-processing.

We give now an analytical expression for the raw key rate. We denote by $P_0$ the probability that the quantum state sent by Alice (Bob) is stored in the quantum memory\footnote{Here, we consider a completely symmetric set-up which implies that the success probability is the same on Alice's and Bob's side. However, in case that Alice and Bob have different probabilities, it is easy to repeat the  analysis keeping these two probabilities different. }.  One knows that this event has happened because the quantum memories are supposed to be heralded. In the following we will measure the time in units of $\Delta t:=\nu_s^{-1}$ which represents the time that the quantum memory has to wait between two attempts. We introduce the probability $P(k_A, k_B)$ that the photons of Alice AND Bob are stored at time-bin $k_A$ and $k_B$ 
and they where not stored before, i.e. 
\begin{equation}
\label{eq:pkakb}
P(k_A, k_B):=P_0^2(1-P_0)^{k_A-1}(1-P_0)^{k_B-1}.
\end{equation}
The average number of attempts by the source necessary for generating one bit of the raw key  is given by 
\begin{align}
\label{eq:avgattemps}
<K>&:=\sum_{s=0}^{\infty}\sum_{k=1}^{\infty}k\cdot s\cdot\nonumber\\
(& P_{BSM}(k|k,k)(1-P_{BSM}(k|k,k))^{s}P(k, k)+\nonumber\\
&+\sum_{i=1}^{k-1}P_{BSM}(k|k,i)(1-P_{BSM}(k|k,i))^{s}P(k, i)\nonumber\\
&+\sum_{i=1}^{k-1}P_{BSM}(k|i,k)(1-P_{BSM}(k|i,k))^{s}P(i, k)),
\end{align}
where $P_{BSM}(k|k_A,k_B)$ is the probability that the BSM was successful at time $k=\mathrm{max}(k_A, k_B)$ when the two involved photons where stored at times $k_A$ and $k_B$. Note that if we consider only the first line containing $P(k,k)$ then we recover the expression for the rate of the relay. The second (third) line accounts for the case that a photon sent by Bob (Alice) has been stored at a certain time $i<k$ and the photon sent by Alice (Bob) has been stored at time $k$. The average time becomes $<T>:=\Delta t <K>$.

\begin{figure}
 \includegraphics[width=.3\textwidth]{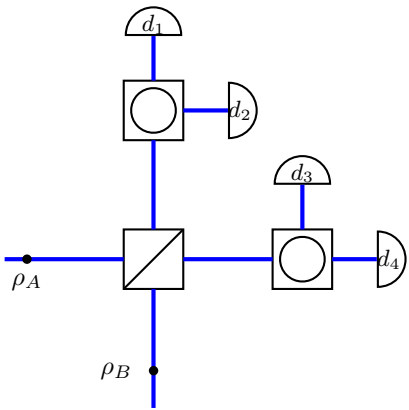}
\caption{(Color online) [adapted from \cite{minar}] Scheme for entanglement swapping with linear optics \cite{0295-5075-25-8-001, sangouard_quantum_2011}. The square with a diagonal line is a polarizing beam splitter in the rectilinear basis and the squares with a circle inside are polarizing beam splitters in the diagonal basis. Entanglement swapping is successful if $d_1$ and $d_3$ click (or $d_1$ and $d_4$ or $d_2$ and $d_3$ or $d_2$ and $d_4$). The state $\rho_{A}$($\rho_{B}$) is produced by Alice(Bob).}
\label{fig:entswap} 
\end{figure}

In order to obtain a closed formula we consider a specific implementation of the BSM \cite{0295-5075-25-8-001, sangouard_quantum_2011} where the photons are first retrieved from the quantum memories and then measured with linear optics (see Fig.~\ref{fig:entswap}). This method is probabilistic and when implemented with perfect quantum memories and detectors leads to a maximal success probability of $\frac{1}{2}$ \cite{calsamiglia2001maximum}. The BSM is successful when a particular two-fold detection happens. We consider practical threshold detectors with detection efficiency $\eta_D$ and dark count probability $p_
D$.  We denote by $\eta_M$ the retrieval probability of a photon from the quantum memory. The BSM success probability for the scheme given in Fig.~\ref{fig:entswap} as a function of $\eta_{MD}:=\eta_M\eta_D$ is then given by \cite{kok_introduction_2010}:
\begin{align}
\label{eq:pesperf}
P_{BSM}(\eta_{MD}):=&\frac{1}{2}(1-p_D)^2(\eta_{MD}^2+2(4-3\eta_{MD})\eta_{MD}p_D\nonumber\\
&+8(1-\eta_{MD})^2p_D^2).
\end{align}
For $p_D=0$ as we expect $P_{BSM}=\frac{\eta_{MD}^2}{2}$. 
Assuming that $\eta_M$ does not depend on the time a simple expression for the average number of attempts in eq.~\eqref{eq:avgattemps} was derived in  \cite{Collins2007, NadjaHybrid},
\begin{equation}
 <K>:=\frac{1}{P_{BSM}(\eta_{MD})}\frac{3-2P_0}{(2-P_0)P_0}.
\end{equation}
In the case of absence of quantum memories we get ${<K>_{\text{relay}}:=\left(P_{BSM}(P_0\eta_{D})\right)^{-1}}$. For small $P_0$ the rate of the repeater scales as $P_0^{-1}$ while the rate for the relay scales as $P_0^{-2}$. Moreover for the repeater, dark counts do not play a role as typically $p_D\ll\eta_{MD}$. The equivalent condition for the relay would be $p_D\ll\eta_{D}P_0$, which is much more difficult to ensure. For the quantum repeater $\eta_M$ plays the role of $P_0$ for the relay.

With the same formalism we calculate the QBER which enters in the formula of the secret key rate. Let $e_j(k|k_A, k_B)$ be the QBER in the basis $j\in\{X,Z\}$ when the BSM has been performed at  time $k$ and the two photons were stored at times $k_A$ and $k_B$, respectively. Then the average QBER in the basis $j$ is given by
\begin{align}
\label{eq:avgqber}
e_j=&\sum_{k=1}^{\infty}\Bigl[e_j(k|k,k) P(k,k)\nonumber\\
&+\sum_{i=1}^{k-1}e_j(k|k,i)P(k,i)\nonumber\\
&+\sum_{i=1}^{k-1}e_j(k|i,k)P(i,k)\Bigr],
\end{align}
where the first line gives the QBER for the case of a quantum relay, i.e. when both photons arrive at the same time. The second and third lines include the contribution to the QBER given by the measurements where one photon arrived at $i<k$ and the second arrives at time $k$. 

Here, we consider a simple model of decoherence where the quantum memory stores perfectly a quantum state for a certain time $\tau$ and then it transforms the quantum state to the identity for $t>\tau$ \cite{Collins2007}. We call $\tau$ the coherence time and measure it in units of $\Delta t$. This model is valid in quantum memories where the fidelity remains approximately constant for a certain time and then it drops very fast. Formally, we have
\begin{align}
\label{eq:qberinf}
e_j(k|k_A, k_B)&:=e_j(\infty)\Theta[\tau-(k-k_A)]\Theta[\tau-(k-k_B)]\nonumber\\
&+\frac{1}{2}(1-\Theta[\tau-(k-k_A)]\Theta[\tau-(k-k_B)]),
\end{align}
where $\Theta[t]$ is the Heaviside step function \cite{abramowitz1964handbook} such that $\Theta[t]=1$ for $t\geq0$ and $\Theta[t]=0$ for $t<0$ and $e_j(\infty)$ is the QBER that would be obtained if the memory does not decohere ($\tau\rightarrow\infty$) and it is given by \cite{abruzzo2012quantum}
\begin{align}
\label{eq:qberperf}
e_X(\infty)&=e_Z(\infty)\nonumber\\
&=\frac{2 p_D \left(2 \left(\eta _{MD}-1\right){}^2 p_D-\left(\eta_{MD}-2\right) \eta _{MD}\right)}{\eta _{MD}^2+8 \left(\eta_{MD}-1\right){}^2 p_D^2+2 \left(4-3 \eta _{MD}\right) \eta _{MD} p_D}.
\end{align}
Inserting eq.~\eqref{eq:pkakb}, eq.~\eqref{eq:qberinf}  and eq.~\eqref{eq:qberperf} in eq.~\eqref{eq:avgqber} we obtain a closed formula for the average QBER:
\begin{equation}
\label{eq:qberdec}
e_j=e_j(\infty)+\frac{1}{2}\frac{\left(\frac{1}{2}-e_j(\infty)\right)(1-P_0)^{1+\tau}}{2-P_0}.
\end{equation}
It is easy to verify $e_j(\infty)\leq e_j \leq \frac{1}{2}$ and moreover $\lim_{\tau\rightarrow\infty}e_j =e_j(\infty)$ and  $\lim_{P_0\rightarrow0}e_j=\frac{1}{2}$. Note that due to our specific set-up $e_X=e_Z$.

\newcommand{\taumin}{\tau^{\textrm MIN}}
If the QBER is too high it is not possible to extract a secret key as the secret key rate in  eq.~\eqref{eq:kr:sps} becomes non-positive. When $e_X=e_Z$ the maximal QBER for a non-zero secret key rate is given by $e^{\text{MAX}}:=0.11$.  A critical parameter is  therefore $\taumin_{SPS}$ which represents the minimal $\tau$ permitting to extract a secret key and can be obtained from eq.~\eqref{eq:qberdec} by requiring that $e_X=e^{\text{MAX}}$. The minimal allowed coherence time  is given by
\begin{equation}
\label{eq:taumin}
 \taumin_{SPS}=\frac{\log \left(\frac{\left(P_0-2\right)
   \left(e_X(\infty)-e^{\text{MAX}}\right)}{\left(P_0-1\right)
   \left(2 e_X(\infty)-1\right)}\right)}{\log \left(1-P_0\right)}.
\end{equation}

In the following section we will provide numbers for the minimal coherence time and the secret key rate in a realistic scenario.

\subsection{Performance}

We discuss now the performance of the protocol as a function of the imperfections of the set-up. Then we analyze the relation with the original \mdirel with single-photon states. We consider an implementation where photons are transmitted through optical fibers. Therefore $P_0:=\eta_T$ where $\eta_T:=10^{-\frac{\alpha L}{2\cdot10}}$ is the probability that a photon has not been absorbed after traveling for a distance $\frac{L}{2}$ and $\alpha$ is the absorption coefficient. Throughout the whole paper we will consider $\alpha=0.17$ dB/km which is the lowest attenuation in common optical fibers. In the following analysis we will consider detectors with detection efficiency $\eta_D=0.2$ and dark count probability $p_D=10^{-6}$. Such detectors are considered optimistic but not unrealistic \cite{Scarani:2009}. Regarding quantum memories we use $\eta_{M}=0.6$ which is a value already achieved experimentally \cite{sangouard_quantum_2011}.

\begin{figure}
 \includegraphics[width=.3\textwidth, angle=-90]{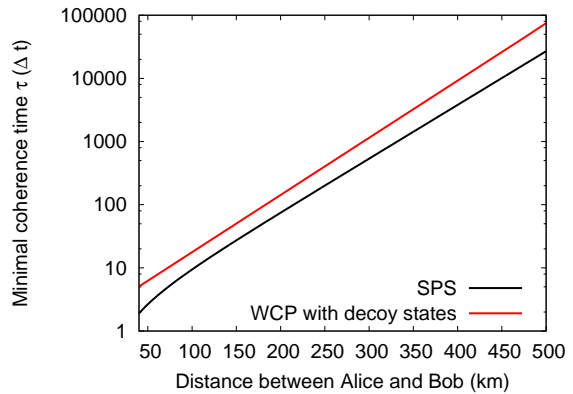}
\caption{(Color online) Minimal coherence time $\taumin$ in units of $\Delta t$ such that the secret key rate is non-zero. Black-solid line: SPS protocol (see eq.~\eqref{eq:taumin}).  Red-solid line: WCP protocol (derived by calculating the zero of eq.~\eqref{eq:rinfinitydecoy}). Parameters:  $\eta_D=0.2$, $\eta_M=0.6$, $p_D=10^{-6}$, $\alpha=0.17$ dB/km.}
\label{fig:lstmintau} 
\end{figure}

In Fig.~\ref{fig:lstmintau} we show $\taumin_{SPS}$ versus the distance between Alice and Bob. For $L=400$ km we get $\taumin\approx4\cdot10^4$ which can be transformed in seconds multiplying by $\Delta t$. For an hypothetical source at $100$ MHz this would correspond to a coherence time of the order of 400 microseconds. Note that single-photon sources at such a speed do not yet exist. We will reconsider this number in the next section when we will consider WCP sources. By increasing the repetition frequency it is possible to use quantum memories with lower coherence times. This is different to standard quantum repeater protocols where the coherence time depends also on the communication time. We see that the curve of $\taumin$ is tightly upper bounded by the average maximal time that is necessary to wait before both quantum memories are filled up. This can be understood by observing 
that for $P_0\ll1$ and $e_X\approx0$ we have $<K>P_{BSM}\approx\frac{3}{2P_0}$ and $\tau^{MIN}\approx\frac{log(2e^{MAX})}{-P_0}\approx\frac{1.51}{P_0}$.   

\begin{figure}
 \includegraphics[width=.3\textwidth, angle=-90]{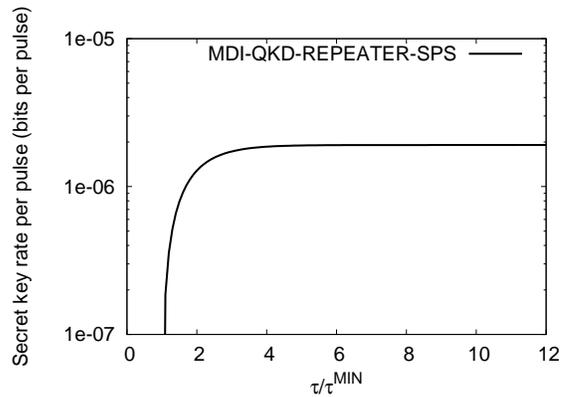}
\caption{(Color online) Secret key rate per pulse as function of $\tau/\taumin_{SPS}$. Parameters: $\eta_D=0.2$, $\eta_M=0.6$, $p_D=10^{-6}$, $\alpha=0.17$ dB/km, $L=400$ km.}
\label{fig:reslst} 
\end{figure}
In Fig.~\ref{fig:reslst} we show the secret key rate as a function of $\tau/\taumin_{SPS}$ for a fixed distance between Alice and Bob ($L=400$ km). We see that a flat region is reached for $\tau\approx 5\taumin_{SPS}$. The same behavior is found also for other values of the distance between Alice and Bob.

\begin{figure}
 \includegraphics[width=.3\textwidth, angle=-90]{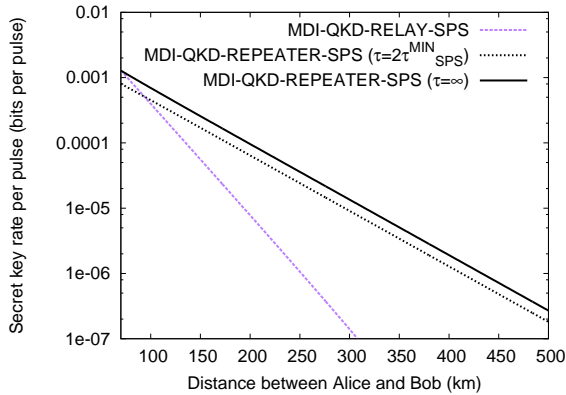}
\caption{(Color online) Secret key rate per pulse versus distance between Alice and Bob. Parameters: $\eta_D=0.2$, $\eta_M=0.6$, $p_D=10^{-6}$, $\alpha=0.17$ dB/km. }
\label{fig:lstkr} 
\end{figure}

Finally, we discuss the secret key rate as a function of the distance and  compare it to a set-up without quantum memories. As shown in Fig.~\ref{fig:lstkr}, the set-up with quantum memories permits to increase significantly the secret key rate with respect to a set-up without quantum memories. For $\etaD=0.2$, $\etaM=0.6$ and $p_D=10^{-6}$ the cross-over distance is around $100$ km. Moreover, we see that the difference between $\tau=2\taumin$ and $\tau=\infty$ is very small. This result suggests  that the protocol is not very susceptible to decoherence of quantum memories: perfect quantum memories are not needed as coherence times slightly bigger than $\taumin$ permit to achieve the maximal secret key rate obtainable with perfect quantum memories. 
   Moreover, we have performed numerical simulations for quantum memories where the decoherence model is depolarization\footnote{The model we have considered is  ${D(\rho):=e^{-\frac{t}{\tau}}\rho+\frac{1-e^{-\frac{t}{\tau}}}{2}\one}$ where $\tau$ is the coherence time.}, and we found that this result does not change qualitatively.

Concluding this section,  we have proven that using single-photon sources and imperfect quantum memories it is possible to essentially double the distance with respect to \mdirel when implemented with single-photon sources. 

\section{Scheme with weak coherent pulse sources\label{sec:wcp}}
A critical assumption of the previous section was that Alice and Bob have on-demand single-photon sources at their disposal.  In this section we consider sources of weak coherent pulses which offer a very high repetition frequency - with current technology even in the order of GHz \cite{5504020}. On the other hand this type of source requires a more complicated security analysis due to the fact that multi-photon pulses are susceptible to the photon-number-splitting (PNS) attacks \cite{PhysRevLett.85.1330}. In order to detect this attack it is possible to use decoy states \cite{PhysRevLett.94.230504,PhysRevA.72.012326}. 
In the scheme with decoy states Alice and Bob prepare phase randomized weak coherent pulses of the form $ \rho=\sum_{n=0}^{\infty}p(n)\ket{n}\bra{n}$ with $p(n):=e^{-\mu}\frac{\mu^{n}}{n!}$. The parameter $\mu$ is the intensity (average photon number) of the pulse.  

The QKD protocol with decoy states \cite{PhysRevLett.94.230504,PhysRevA.72.012326} which we employ here is analogous to the one described in sec.~\ref{sec:sps}, apart from the following differences:
\begin{itemize}
 \item when Alice and Bob prepare the state they choose at random and independently also its intensity $\mu$ which is a continuous parameter with $0\leq\mu<\infty$. One particular intensity $\overline{\mu}$ is chosen with probability almost one,
\item the measurements for pulses with intensity $\overline{\mu}$ are used for extracting a secret key, whereas the others are used for detecting Eve's PNS attack. 
\end{itemize}

The formula for the secret key rate is analogous to eq.~\eqref{eq:kr:sps} with the modifications due to the fact that Eve can perform  PNS attacks. It is given by \cite{PhysRevLett.108.130503}:
\begin{equation}
\label{eq:rinfinitydecoy}
r_{\infty}:=\max_{\mu>0}\left[\frac{1}{<T>}(f_{11}(1-h(e_{X}^{11}))-h(e_Z))\right],
\end{equation}
where $f_{11}$ is the fraction of bits in the raw key which are generated when Alice and Bob send single-photon states and $e_X^{11}$ is the QBER of these bits. The QBER $e_X^{11}$, is accessible due to the fact that we use decoy states \cite{PhysRevLett.108.130503}. The QBER $e_Z$ is determined using all data.  All  quantities entering in the formula of the secret key rate in eq.~\eqref{eq:rinfinitydecoy} depend on a generic intensity $\mu$ which is then maximized. The optimal intensity is denoted by $\overline{\mu}$ (see above).  In the following we derive analytical expressions for these parameters as function of the imperfections of the set-up. We will assume that detectors have no dark counts. This will permit to have closed formulas which will allow to understand the role of each parameter. Dark counts do not play a crucial role as long as  $\eta_{MD}\gg p_D$. For realistic choice of parameters this condition is easily satisfied. 

Given a pulse of $n$-photons, the probability that at least one photon is stored into the quantum memory is given by $(1-(1-\eta_{T})^n)$ where $\eta_{T}$ is the probability that one photon has not been absorbed by the quantum channel. In general, the probability $P_0$ that a state has been stored into the quantum memory is given by
\begin{align}
 P_0&:=\sum_{n=1}^{\infty}p(n)(1-(1-\eta_{T})^n)\nonumber\\
 &=1-e^{-\mu\eta_T},
\end{align}
which for $\mu\eta_{T}\ll1$ reduces to $P_0=\mu\eta_{T}$ as expected.

The BSM success probability depends on the probability to store a state with $n$-photons given that the source has generated a state of $m$-photons with $m\geq n$. Formally,
\begin{align}
P(n)&:=\sum_{m=n}^{\infty}p(m)\binom{m}{n}\eta_T^{n}(1-\eta_T)^{m-n}\nonumber\\
&=\frac{(\eta_T\mu)^{n}}{n!}e^{-\eta_T\mu}.
\end{align}
The quantity $\binom{m}{n}\eta_T^{n}(1-\eta_T)^{m-n}$ is the probability that $n$ photons survive from a state with $m$-photons after the transmission through the channel. The probability that the BSM is successful given that one quantum memory contains $n_a$ photons and the other $n_b$ photons is given by (see the appendix for our derivation)
\begin{align}
\label{eq:bsmnanb}
 P_{BSM}(n_a, n_b)&=[(1-\frac{\eta_{MD}}{2})^{n_a} - (1-\eta_{MD})^{n_a}]\cdot\nonumber\\
 &\cdot[(1-\frac{\eta_{MD}}{2})^{n_b}-(1-\eta_{MD})^{n_b}].
\end{align}
For $n_a=n_b=1$ we obtain $P_{BSM}(1, 1)=\frac{1}{2}\eta_{MD}^2$ in accordance to eq.~\eqref{eq:pesperf}. Thus, the BSM success probability is given by
\begin{subequations}
\begin{align}
 P_{BSM}&:=2\frac{\sum_{n_a=1}^{\infty}\sum_{n_b=1}^{\infty}P(n_a)P(n_b)P_{BSM}(n_a, n_b)}{\sum_{n_a=1}^{\infty}\sum_{n_b=1}^{\infty}P(n_a)P(n_b)}\label{eq:pbsmdecoy1}\\
 &=2\frac{e^{-2 \mu  \eta _T \left(\eta _{MD} -1\right)}
   \left(e^{\frac{1}{2} \mu  \eta _{MD}  \eta
   _T}-1\right){}^2}{\left(e^{\mu  \eta _T}-1\right){}^2}.
\end{align}
\end{subequations}
The denominator in eq.~\eqref{eq:pbsmdecoy1} gives the probability that two photons are stored in the quantum memories which is equal to $P_0^2$. The numerator is the total probability of all  events in which the BSM is successful when one quantum memory contains $n_a$ photons and the other one contains $n_b$ photons. The factor $2$ comes from the fact that the BSM with linear optics can distinguish only two Bell states. For the limiting case $\mu\eta_T\ll1$ we obtain $P_{BSM}=P_{BSM}(1,1)$.

The fraction of measurements coming from single-photons is denoted as $f_{11}$ and given by 
\begin{subequations}
\begin{align}
 f_{11}&=\frac{P(1)^2 P_{BSM}(1,1)}{\sum_{n_a=1}^{\infty}\sum_{n_b=1}^{\infty}P(n_a)P(n_b)}\label{eq:f11decoy}\\
 &=\frac{\mu ^2 \eta _{MD}^2  \eta _T^2 e^{\mu  \eta_{MD}  \eta _T-2
   \mu }}{4 \left(e^{\frac{1}{2} \mu  \eta _{MD}  \eta _T}-1\right){}^2},
\end{align}
\end{subequations}
which in the limit $\mu\eta_T\ll1$ becomes $f_{11}=1$ as in this limit all measurements come from single-photon states. The numerator of eq.~\eqref{eq:f11decoy} represents the probability that the sources of Alice and Bob produce  single-photons which are stored in the quantum memories and which lead to successful BSM. The denominator is the total probability to obtain a state which does not contain the vacuum.

 Regarding the QBER we observe that if there are no dark counts then both $e_{X}^{11}$ and $e_{Z}$ are zero. This property of the protocol has been discussed also in \cite{PhysRevLett.108.130503}. Therefore, errors will arise only due to decoherence. The calculation is analogous to the one for  single-photon sources of sec.~\ref{sec:scrspsa}. We assume the same decoherence model. The only difference comes from the fact that now $P_0$ is different, in particular we have
\begin{align}
 e_{X}^{11}&=e_X^{11}(\infty)+\frac{1}{2}\frac{\left(\frac{1}{2}-e_X^{11}(\infty)\right)(1-P_0^{11})^{1+\tau}}{2-P_0^{11}},\\
 e_{Z}&=e_Z(\infty)+\frac{1}{2}\frac{\left(\frac{1}{2}-e_Z(\infty)\right)(1-P_0)^{1+\tau}}{2-P_0},
\end{align}
with $P_{0}^{11}=p(1)\eta_{T}$ the probability to store single-photon states in one quantum memory.  

We have thus derived all quantities present in the formula of the secret key rate, and we can now evaluate and characterize the protocol.

\begin{figure}
 \includegraphics[width=.3\textwidth, angle=-90]{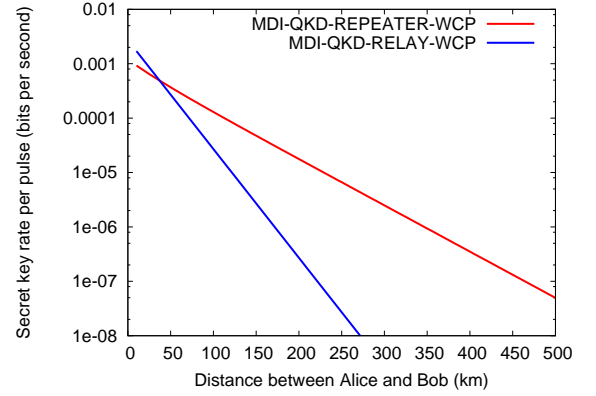}
\caption{(Color online) Secret key rate versus distance between Alice and Bob. Comparison between relay \cite{PhysRevLett.108.130503} (blue) and repeater (see eq.~\eqref{eq:rinfinitydecoy} )(red). Parameters: $\eta_D=0.2$, $\eta_M=0.6$, $p_D=0$, $\alpha=0.17$ dB/km, $\tau=\infty$.}
\label{fig:krWCP} 
\end{figure}

In Fig.~\ref{fig:krWCP} we show the comparison between \mdirepw and \mdirelw. As we see quantum memories permit to increase significantly the secret key rate or the distance where it is possible to perform QKD.

\vspace{.5cm}

As shown in Fig.~\ref{fig:lstmintau}, the minimally allowed  coherence time $\tau^{MIN}_{WCP}$ is larger then $\tau^{MIN}_{SPS}$. The reason is that now the produced state contains also a vacuum that reduces the probability that a photon arrives to the quantum memory. However, the difference is  less than one order of magnitude. Moreover, analogously to the case of SPS the flat region($\tau\rightarrow\infty$) of the secret key rate is reached already with $\tau=5\tau^{MIN}_{WCP}$.

In practical cases, only a finite number of different decoy states is used. In order to adapt our result to this case it is enough to use the results of \cite{PhysRevA.86.062319}. Moreover, finite-size corrections are necessary for giving realistic estimates. This can be done by adopting  the formalism developed in \cite{PhysRevA.86.022332, PhysRevA.86.052305, 2013arXiv1305.6965X}

\section{Conclusions\label{sec:conc}}

In this paper we have explored the possibility to enable long distance QKD without entanglement sources. We have shown that when quantum memories are used it is possible to improve the distance where measurement-device-independent quantum key distribution can be implemented. Moreover, we have shown that the protocol we consider in this paper is robust against common device imperfections such as detector efficiency, quantum memory retrieval efficiency and finite decoherence time. We believe that our result could be used as a first step in the development of long-distance quantum key distribution.  It requires weak coherent pulse sources, which are already available commercially, and heralded quantum memories which are under current development. 

\section*{Acknowledgments}
We would like to thank Sylvia Bratzik and Tobias Moroder for valuable and enlightening discussions. We acknowledge financial support by the German Federal Ministry of Education and Research (BMBF, Project QuOReP). 

\section*{APPENDIX}
We prove eq.~\eqref{eq:bsmnanb} when the Bell-state measurement is done between two WCP states  in the computational basis. The proof for the case of the diagonal basis is analogous.

We define
\newcommand{\Gtr}[6]{G_{#1 #2 #3 #4}\left(#5, #6\right)}
\newcommand{\Gtrdef}[6]{\mathrm{tr}\left(\Pi_{d_{#1}}^{(1)}\Pi_{d_{#2}}^{(0)}\Pi_{d_{#3}}^{(1)}\Pi_{d_{#4}}^{(0)}\mathcal{E}\left(#5\otimes #6 \right)\right)}
\begin{widetext}
\begin{equation}
\Gtr{i_1}{i_2}{i_3}{i_4}{\rho_{A}^{(n_a)}}{\rho_{B}^{(n_b)}}:=\Gtrdef{i_1}{i_2}{i_3}{i_4}{\rho_{A}^{(n_a)}}{\rho_{B}^{(n_b)}}
\end{equation}
\end{widetext}
where $\mathcal{E}$ represents the action of the partial BSM and is given by the following mapping (see Fig.~\ref{fig:entswap})
\begin{align}
 b_{H}\rightarrow \frac{d_3+d_4}{2}&, \quad b_{V}\rightarrow \frac{d_1-d_2}{2},\\
 a_{H}\rightarrow \frac{d_1+d_2}{2}&, \quad a_{V}\rightarrow \frac{d_3-d_4}{2},
\end{align}
where $a_H, a_V$ are the modes of $\rho_A$ and $b_H, b_V$ are the modes of $\rho_B$. The POVM elements of threshold detectors are given by
\begin{align}
 \Pi^{(0)}:=\sum_{i=0}^{\infty}(1-\eta_D)^i\ket{i}\bra{i},
  \Pi^{(1)}:=\sum_{i=0}^{\infty}(1-(1-\eta_D)^i)\ket{i}\bra{i}.
\end{align}
The success probability of a BSM is given by
\begin{widetext}
\begin{equation}
 P_{BSM}(n_a, n_b):=\frac{1}{4}\sum_{i_1 i_2 i_3 i_4\in\mathcal{A}}\sum_{\phi\in\mathcal{B}}\Gtr{i_1}{i_2}{i_3}{i_4}{\phi^{\otimes n_a}}{\phi^{\otimes n_b}},
\end{equation}
\end{widetext}
where $\mathcal{A}=\{1234, 1243, 2134, 2143\}$ is the set containing the combinations of two-fold detection leading to a successful entanglement swapping and $\mathcal{B}=\{\ket{HH}\bra{HH}, \ket{VV}\bra{VV}\}$ is a set containing the quantum states produced by the two sources of Alice and Bob when they choose the computational basis. The set $\mathcal{B}$ does not contain the cross-terms like $\sigma:=\ket{HH}\bra{VV}$ because $\Gtr{i_1}{i_2}{i_3}{i_4}{\sigma^{\otimes n_a}}{\sigma^{\otimes n_b}}=0$. Due to the symmetries of the map $\mathcal{E}$ we find that the function G is equal for all combinations of indices in $\mathcal{A}$ and quantum states in $\mathcal{B}$, therefore
\begin{widetext}
\begin{equation}
 P_{BSM}(n_a, n_b)=\frac{4\cdot2}{4}\Gtr{1}{2}{3}{4}{\ket{HH}\bra{HH}^{\otimes n_a}}{\ket{HH}\bra{HH}^{\otimes n_b}}.
\end{equation}
\end{widetext}
Using the fact that $\ket{HH}:=a_{H}^{\dagger}b_{H}^{\dagger}\ket{0}$ and using the definition of $\mathcal{E}$ it is straightforward, but lengthly, to calculate $G$ and finally to find the result in eq.~\eqref{eq:bsmnanb}.

\bibliography{draft}

\begin{thebibliography}{32}%
\makeatletter
\providecommand \@ifxundefined [1]{%
 \@ifx{#1\undefined}
}%
\providecommand \@ifnum [1]{%
 \ifnum #1\expandafter \@firstoftwo
 \else \expandafter \@secondoftwo
 \fi
}%
\providecommand \@ifx [1]{%
 \ifx #1\expandafter \@firstoftwo
 \else \expandafter \@secondoftwo
 \fi
}%
\providecommand \natexlab [1]{#1}%
\providecommand \enquote  [1]{``#1''}%
\providecommand \bibnamefont  [1]{#1}%
\providecommand \bibfnamefont [1]{#1}%
\providecommand \citenamefont [1]{#1}%
\providecommand \href@noop [0]{\@secondoftwo}%
\providecommand \href [0]{\begingroup \@sanitize@url \@href}%
\providecommand \@href[1]{\@@startlink{#1}\@@href}%
\providecommand \@@href[1]{\endgroup#1\@@endlink}%
\providecommand \@sanitize@url [0]{\catcode `\\12\catcode `\$12\catcode
  `\&12\catcode `\#12\catcode `\^12\catcode `\_12\catcode `\%12\relax}%
\providecommand \@@startlink[1]{}%
\providecommand \@@endlink[0]{}%
\providecommand \url  [0]{\begingroup\@sanitize@url \@url }%
\providecommand \@url [1]{\endgroup\@href {#1}{\urlprefix }}%
\providecommand \urlprefix  [0]{URL }%
\providecommand \Eprint [0]{\href }%
\providecommand \doibase [0]{http://dx.doi.org/}%
\providecommand \selectlanguage [0]{\@gobble}%
\providecommand \bibinfo  [0]{\@secondoftwo}%
\providecommand \bibfield  [0]{\@secondoftwo}%
\providecommand \translation [1]{[#1]}%
\providecommand \BibitemOpen [0]{}%
\providecommand \bibitemStop [0]{}%
\providecommand \bibitemNoStop [0]{.\EOS\space}%
\providecommand \EOS [0]{\spacefactor3000\relax}%
\providecommand \BibitemShut  [1]{\csname bibitem#1\endcsname}%
\let\auto@bib@innerbib\@empty
\bibitem [{\citenamefont {Scarani}\ \emph {et~al.}(2009)\citenamefont
  {Scarani}, \citenamefont {Bechmann-Pasquinucci}, \citenamefont {Cerf},
  \citenamefont {Du\ifmmode~\check{s}\else \v{s}\fi{}ek}, \citenamefont
  {L\"utkenhaus},\ and\ \citenamefont {Peev}}]{Scarani:2009}%
  \BibitemOpen
  \bibfield  {author} {\bibinfo {author} {\bibfnamefont {V.}~\bibnamefont
  {Scarani}}, \bibinfo {author} {\bibfnamefont {H.}~\bibnamefont
  {Bechmann-Pasquinucci}}, \bibinfo {author} {\bibfnamefont {N.~J.}\
  \bibnamefont {Cerf}}, \bibinfo {author} {\bibfnamefont {M.}~\bibnamefont
  {Du\ifmmode~\check{s}\else \v{s}\fi{}ek}}, \bibinfo {author} {\bibfnamefont
  {N.}~\bibnamefont {L\"utkenhaus}}, \ and\ \bibinfo {author} {\bibfnamefont
  {M.}~\bibnamefont {Peev}},\ }\href {\doibase 10.1103/RevModPhys.81.1301}
  {\bibfield  {journal} {\bibinfo  {journal} {Rev. Mod. Phys.}\ }\textbf
  {\bibinfo {volume} {81}},\ \bibinfo {pages} {1301} (\bibinfo {year}
  {2009})}\BibitemShut {NoStop}%
\bibitem [{\citenamefont {Briegel}\ \emph {et~al.}(1998)\citenamefont
  {Briegel}, \citenamefont {D\"{u}r}, \citenamefont {Cirac},\ and\
  \citenamefont {Zoller}}]{briegel_quantum_1998}%
  \BibitemOpen
  \bibfield  {author} {\bibinfo {author} {\bibfnamefont {H.~J.}\ \bibnamefont
  {Briegel}}, \bibinfo {author} {\bibfnamefont {W.}~\bibnamefont {D\"{u}r}},
  \bibinfo {author} {\bibfnamefont {J.~I.}\ \bibnamefont {Cirac}}, \ and\
  \bibinfo {author} {\bibfnamefont {P.}~\bibnamefont {Zoller}},\ }\href
  {\doibase 10.1103/PhysRevLett.81.5932} {\bibfield  {journal} {\bibinfo
  {journal} {Phys. Rev. Lett.}\ }\textbf {\bibinfo {volume} {81}},\ \bibinfo
  {pages} {5932{\textendash}5935} (\bibinfo {year} {1998})}\BibitemShut
  {NoStop}%
\bibitem [{\citenamefont {Sangouard}\ \emph {et~al.}(2011)\citenamefont
  {Sangouard}, \citenamefont {Simon}, \citenamefont {de~Riedmatten},\ and\
  \citenamefont {Gisin}}]{sangouard_quantum_2011}%
  \BibitemOpen
  \bibfield  {author} {\bibinfo {author} {\bibfnamefont {N.}~\bibnamefont
  {Sangouard}}, \bibinfo {author} {\bibfnamefont {C.}~\bibnamefont {Simon}},
  \bibinfo {author} {\bibfnamefont {H.}~\bibnamefont {de~Riedmatten}}, \ and\
  \bibinfo {author} {\bibfnamefont {N.}~\bibnamefont {Gisin}},\ }\href
  {\doibase 10.1103/RevModPhys.83.33} {\bibfield  {journal} {\bibinfo
  {journal} {Reviews of Modern Physics}\ }\textbf {\bibinfo {volume} {83}},\
  \bibinfo {pages} {33} (\bibinfo {year} {2011})}\BibitemShut {NoStop}%
\bibitem [{\citenamefont {Sangouard}(2012)}]{sangouard2012quantum}%
  \BibitemOpen
  \bibfield  {author} {\bibinfo {author} {\bibfnamefont {N.}~\bibnamefont
  {Sangouard}},\ }\href@noop {} {\bibfield  {journal} {\bibinfo  {journal}
  {Nature Photonics}\ }\textbf {\bibinfo {volume} {6}},\ \bibinfo {pages} {722}
  (\bibinfo {year} {2012})}\BibitemShut {NoStop}%
\bibitem [{\citenamefont {Lo}\ \emph {et~al.}(2012)\citenamefont {Lo},
  \citenamefont {Curty},\ and\ \citenamefont {Qi}}]{PhysRevLett.108.130503}%
  \BibitemOpen
  \bibfield  {author} {\bibinfo {author} {\bibfnamefont {H.-K.}\ \bibnamefont
  {Lo}}, \bibinfo {author} {\bibfnamefont {M.}~\bibnamefont {Curty}}, \ and\
  \bibinfo {author} {\bibfnamefont {B.}~\bibnamefont {Qi}},\ }\href {\doibase
  10.1103/PhysRevLett.108.130503} {\bibfield  {journal} {\bibinfo  {journal}
  {Phys. Rev. Lett.}\ }\textbf {\bibinfo {volume} {108}},\ \bibinfo {pages}
  {130503} (\bibinfo {year} {2012})}\BibitemShut {NoStop}%
\bibitem [{\citenamefont {Braunstein}\ and\ \citenamefont
  {Pirandola}(2012)}]{PhysRevLett.108.130502}%
  \BibitemOpen
  \bibfield  {author} {\bibinfo {author} {\bibfnamefont {S.~L.}\ \bibnamefont
  {Braunstein}}\ and\ \bibinfo {author} {\bibfnamefont {S.}~\bibnamefont
  {Pirandola}},\ }\href {\doibase 10.1103/PhysRevLett.108.130502} {\bibfield
  {journal} {\bibinfo  {journal} {Phys. Rev. Lett.}\ }\textbf {\bibinfo
  {volume} {108}},\ \bibinfo {pages} {130502} (\bibinfo {year}
  {2012})}\BibitemShut {NoStop}%
\bibitem [{\citenamefont {de~Riedmatten}\ \emph {et~al.}(2004)\citenamefont
  {de~Riedmatten}, \citenamefont {Marcikic}, \citenamefont {Tittel},
  \citenamefont {Zbinden}, \citenamefont {Collins},\ and\ \citenamefont
  {Gisin}}]{PhysRevLett.92.047904}%
  \BibitemOpen
  \bibfield  {author} {\bibinfo {author} {\bibfnamefont {H.}~\bibnamefont
  {de~Riedmatten}}, \bibinfo {author} {\bibfnamefont {I.}~\bibnamefont
  {Marcikic}}, \bibinfo {author} {\bibfnamefont {W.}~\bibnamefont {Tittel}},
  \bibinfo {author} {\bibfnamefont {H.}~\bibnamefont {Zbinden}}, \bibinfo
  {author} {\bibfnamefont {D.}~\bibnamefont {Collins}}, \ and\ \bibinfo
  {author} {\bibfnamefont {N.}~\bibnamefont {Gisin}},\ }\href {\doibase
  10.1103/PhysRevLett.92.047904} {\bibfield  {journal} {\bibinfo  {journal}
  {Phys. Rev. Lett.}\ }\textbf {\bibinfo {volume} {92}},\ \bibinfo {pages}
  {047904} (\bibinfo {year} {2004})}\BibitemShut {NoStop}%
\bibitem [{\citenamefont {Lydersen}\ \emph {et~al.}(2010)\citenamefont
  {Lydersen}, \citenamefont {Wiechers}, \citenamefont {Wittmann}, \citenamefont
  {Elser}, \citenamefont {Skaar},\ and\ \citenamefont
  {Makarov}}]{lydersen2010hacking}%
  \BibitemOpen
  \bibfield  {author} {\bibinfo {author} {\bibfnamefont {L.}~\bibnamefont
  {Lydersen}}, \bibinfo {author} {\bibfnamefont {C.}~\bibnamefont {Wiechers}},
  \bibinfo {author} {\bibfnamefont {C.}~\bibnamefont {Wittmann}}, \bibinfo
  {author} {\bibfnamefont {D.}~\bibnamefont {Elser}}, \bibinfo {author}
  {\bibfnamefont {J.}~\bibnamefont {Skaar}}, \ and\ \bibinfo {author}
  {\bibfnamefont {V.}~\bibnamefont {Makarov}},\ }\href@noop {} {\bibfield
  {journal} {\bibinfo  {journal} {Nature photonics}\ }\textbf {\bibinfo
  {volume} {4}},\ \bibinfo {pages} {686} (\bibinfo {year} {2010})}\BibitemShut
  {NoStop}%
\bibitem [{\citenamefont {Gerhardt}\ \emph {et~al.}(2011)\citenamefont
  {Gerhardt}, \citenamefont {Liu}, \citenamefont {Lamas-Linares}, \citenamefont
  {Skaar}, \citenamefont {Kurtsiefer},\ and\ \citenamefont
  {Makarov}}]{gerhardt2011full}%
  \BibitemOpen
  \bibfield  {author} {\bibinfo {author} {\bibfnamefont {I.}~\bibnamefont
  {Gerhardt}}, \bibinfo {author} {\bibfnamefont {Q.}~\bibnamefont {Liu}},
  \bibinfo {author} {\bibfnamefont {A.}~\bibnamefont {Lamas-Linares}}, \bibinfo
  {author} {\bibfnamefont {J.}~\bibnamefont {Skaar}}, \bibinfo {author}
  {\bibfnamefont {C.}~\bibnamefont {Kurtsiefer}}, \ and\ \bibinfo {author}
  {\bibfnamefont {V.}~\bibnamefont {Makarov}},\ }\href@noop {} {\bibfield
  {journal} {\bibinfo  {journal} {Nature Communications}\ }\textbf {\bibinfo
  {volume} {2}},\ \bibinfo {pages} {349} (\bibinfo {year} {2011})}\BibitemShut
  {NoStop}%
\bibitem [{\citenamefont {{Rubenok}}\ \emph {et~al.}(2012)\citenamefont
  {{Rubenok}}, \citenamefont {{Slater}}, \citenamefont {{Chan}}, \citenamefont
  {{Lucio-Martinez}},\ and\ \citenamefont {{Tittel}}}]{2012arXiv1204.0738R}%
  \BibitemOpen
  \bibfield  {author} {\bibinfo {author} {\bibfnamefont {A.}~\bibnamefont
  {{Rubenok}}}, \bibinfo {author} {\bibfnamefont {J.~A.}\ \bibnamefont
  {{Slater}}}, \bibinfo {author} {\bibfnamefont {P.}~\bibnamefont {{Chan}}},
  \bibinfo {author} {\bibfnamefont {I.}~\bibnamefont {{Lucio-Martinez}}}, \
  and\ \bibinfo {author} {\bibfnamefont {W.}~\bibnamefont {{Tittel}}},\
  }\href@noop {} {\bibfield  {journal} {\bibinfo  {journal} {ArXiv e-prints}\ }
  (\bibinfo {year} {2012})},\ \Eprint {http://arxiv.org/abs/1204.0738}
  {arXiv:1204.0738 [quant-ph]} \BibitemShut {NoStop}%
\bibitem [{\citenamefont {{Ferreira da Silva}}\ \emph
  {et~al.}(2012)\citenamefont {{Ferreira da Silva}}, \citenamefont
  {{Vitoreti}}, \citenamefont {{Xavier}}, \citenamefont {{Tempor{\~a}o}},\ and\
  \citenamefont {{von der Weid}}}]{2012arXiv1207.6345F}%
  \BibitemOpen
  \bibfield  {author} {\bibinfo {author} {\bibfnamefont {T.}~\bibnamefont
  {{Ferreira da Silva}}}, \bibinfo {author} {\bibfnamefont {D.}~\bibnamefont
  {{Vitoreti}}}, \bibinfo {author} {\bibfnamefont {G.~B.}\ \bibnamefont
  {{Xavier}}}, \bibinfo {author} {\bibfnamefont {G.~P.}\ \bibnamefont
  {{Tempor{\~a}o}}}, \ and\ \bibinfo {author} {\bibfnamefont {J.~P.}\
  \bibnamefont {{von der Weid}}},\ }\href@noop {} {\bibfield  {journal}
  {\bibinfo  {journal} {ArXiv e-prints}\ } (\bibinfo {year} {2012})},\ \Eprint
  {http://arxiv.org/abs/1207.6345} {arXiv:1207.6345 [quant-ph]} \BibitemShut
  {NoStop}%
\bibitem [{\citenamefont {{Liu}}\ \emph {et~al.}(2012)\citenamefont {{Liu}},
  \citenamefont {{Chen}}, \citenamefont {{Wang}}, \citenamefont {{Liang}},
  \citenamefont {{Shentu}}, \citenamefont {{Wang}}, \citenamefont {{Cui}},
  \citenamefont {{Yin}}, \citenamefont {{Liu}}, \citenamefont {{Li}},
  \citenamefont {{Ma}}, \citenamefont {{Pelc}}, \citenamefont {{Fejer}},
  \citenamefont {{Zhang}},\ and\ \citenamefont {{Pan}}}]{2012arXiv1209.6178L}%
  \BibitemOpen
  \bibfield  {author} {\bibinfo {author} {\bibfnamefont {Y.}~\bibnamefont
  {{Liu}}}, \bibinfo {author} {\bibfnamefont {T.-Y.}\ \bibnamefont {{Chen}}},
  \bibinfo {author} {\bibfnamefont {L.-J.}\ \bibnamefont {{Wang}}}, \bibinfo
  {author} {\bibfnamefont {H.}~\bibnamefont {{Liang}}}, \bibinfo {author}
  {\bibfnamefont {G.-L.}\ \bibnamefont {{Shentu}}}, \bibinfo {author}
  {\bibfnamefont {J.}~\bibnamefont {{Wang}}}, \bibinfo {author} {\bibfnamefont
  {K.}~\bibnamefont {{Cui}}}, \bibinfo {author} {\bibfnamefont {H.-L.}\
  \bibnamefont {{Yin}}}, \bibinfo {author} {\bibfnamefont {N.-L.}\ \bibnamefont
  {{Liu}}}, \bibinfo {author} {\bibfnamefont {L.}~\bibnamefont {{Li}}},
  \bibinfo {author} {\bibfnamefont {X.}~\bibnamefont {{Ma}}}, \bibinfo {author}
  {\bibfnamefont {J.~S.}\ \bibnamefont {{Pelc}}}, \bibinfo {author}
  {\bibfnamefont {M.~M.}\ \bibnamefont {{Fejer}}}, \bibinfo {author}
  {\bibfnamefont {Q.}~\bibnamefont {{Zhang}}}, \ and\ \bibinfo {author}
  {\bibfnamefont {J.-W.}\ \bibnamefont {{Pan}}},\ }\href@noop {} {\bibfield
  {journal} {\bibinfo  {journal} {ArXiv e-prints}\ } (\bibinfo {year}
  {2012})},\ \Eprint {http://arxiv.org/abs/1209.6178} {arXiv:1209.6178
  [quant-ph]} \BibitemShut {NoStop}%
\bibitem [{\citenamefont {Ma}\ and\ \citenamefont
  {Razavi}(2012)}]{PhysRevA.86.062319}%
  \BibitemOpen
  \bibfield  {author} {\bibinfo {author} {\bibfnamefont {X.}~\bibnamefont
  {Ma}}\ and\ \bibinfo {author} {\bibfnamefont {M.}~\bibnamefont {Razavi}},\
  }\href {\doibase 10.1103/PhysRevA.86.062319} {\bibfield  {journal} {\bibinfo
  {journal} {Phys. Rev. A}\ }\textbf {\bibinfo {volume} {86}},\ \bibinfo
  {pages} {062319} (\bibinfo {year} {2012})}\BibitemShut {NoStop}%
\bibitem [{\citenamefont {Tamaki}\ \emph {et~al.}(2012)\citenamefont {Tamaki},
  \citenamefont {Lo}, \citenamefont {Fung},\ and\ \citenamefont
  {Qi}}]{PhysRevA.85.042307}%
  \BibitemOpen
  \bibfield  {author} {\bibinfo {author} {\bibfnamefont {K.}~\bibnamefont
  {Tamaki}}, \bibinfo {author} {\bibfnamefont {H.-K.}\ \bibnamefont {Lo}},
  \bibinfo {author} {\bibfnamefont {C.-H.~F.}\ \bibnamefont {Fung}}, \ and\
  \bibinfo {author} {\bibfnamefont {B.}~\bibnamefont {Qi}},\ }\href {\doibase
  10.1103/PhysRevA.85.042307} {\bibfield  {journal} {\bibinfo  {journal} {Phys.
  Rev. A}\ }\textbf {\bibinfo {volume} {85}},\ \bibinfo {pages} {042307}
  (\bibinfo {year} {2012})}\BibitemShut {NoStop}%
\bibitem [{\citenamefont {{Xu}}\ \emph {et~al.}(2013)\citenamefont {{Xu}},
  \citenamefont {{Curty}}, \citenamefont {{Qi}},\ and\ \citenamefont
  {{Lo}}}]{2013arXiv1305.6965X}%
  \BibitemOpen
  \bibfield  {author} {\bibinfo {author} {\bibfnamefont {F.}~\bibnamefont
  {{Xu}}}, \bibinfo {author} {\bibfnamefont {M.}~\bibnamefont {{Curty}}},
  \bibinfo {author} {\bibfnamefont {B.}~\bibnamefont {{Qi}}}, \ and\ \bibinfo
  {author} {\bibfnamefont {H.-K.}\ \bibnamefont {{Lo}}},\ }\href@noop {}
  {\bibfield  {journal} {\bibinfo  {journal} {ArXiv e-prints}\ } (\bibinfo
  {year} {2013})},\ \Eprint {http://arxiv.org/abs/1305.6965} {arXiv:1305.6965
  [quant-ph]} \BibitemShut {NoStop}%
\bibitem [{\citenamefont {Song}\ \emph {et~al.}(2012)\citenamefont {Song},
  \citenamefont {Wen}, \citenamefont {Guo},\ and\ \citenamefont
  {Tan}}]{PhysRevA.86.022332}%
  \BibitemOpen
  \bibfield  {author} {\bibinfo {author} {\bibfnamefont {T.-T.}\ \bibnamefont
  {Song}}, \bibinfo {author} {\bibfnamefont {Q.-Y.}\ \bibnamefont {Wen}},
  \bibinfo {author} {\bibfnamefont {F.-Z.}\ \bibnamefont {Guo}}, \ and\
  \bibinfo {author} {\bibfnamefont {X.-Q.}\ \bibnamefont {Tan}},\ }\href
  {\doibase 10.1103/PhysRevA.86.022332} {\bibfield  {journal} {\bibinfo
  {journal} {Phys. Rev. A}\ }\textbf {\bibinfo {volume} {86}},\ \bibinfo
  {pages} {022332} (\bibinfo {year} {2012})}\BibitemShut {NoStop}%
\bibitem [{\citenamefont {Ma}\ \emph {et~al.}(2012)\citenamefont {Ma},
  \citenamefont {Fung},\ and\ \citenamefont {Razavi}}]{PhysRevA.86.052305}%
  \BibitemOpen
  \bibfield  {author} {\bibinfo {author} {\bibfnamefont {X.}~\bibnamefont
  {Ma}}, \bibinfo {author} {\bibfnamefont {C.-H.~F.}\ \bibnamefont {Fung}}, \
  and\ \bibinfo {author} {\bibfnamefont {M.}~\bibnamefont {Razavi}},\ }\href
  {\doibase 10.1103/PhysRevA.86.052305} {\bibfield  {journal} {\bibinfo
  {journal} {Phys. Rev. A}\ }\textbf {\bibinfo {volume} {86}},\ \bibinfo
  {pages} {052305} (\bibinfo {year} {2012})}\BibitemShut {NoStop}%
\bibitem [{\citenamefont {Lo}\ \emph {et~al.}(2005{\natexlab{a}})\citenamefont
  {Lo}, \citenamefont {Ma},\ and\ \citenamefont
  {Chen}}]{PhysRevLett.94.230504}%
  \BibitemOpen
  \bibfield  {author} {\bibinfo {author} {\bibfnamefont {H.-K.}\ \bibnamefont
  {Lo}}, \bibinfo {author} {\bibfnamefont {X.}~\bibnamefont {Ma}}, \ and\
  \bibinfo {author} {\bibfnamefont {K.}~\bibnamefont {Chen}},\ }\href {\doibase
  10.1103/PhysRevLett.94.230504} {\bibfield  {journal} {\bibinfo  {journal}
  {Phys. Rev. Lett.}\ }\textbf {\bibinfo {volume} {94}},\ \bibinfo {pages}
  {230504} (\bibinfo {year} {2005}{\natexlab{a}})}\BibitemShut {NoStop}%
\bibitem [{\citenamefont {Ma}\ \emph {et~al.}(2005)\citenamefont {Ma},
  \citenamefont {Qi}, \citenamefont {Zhao},\ and\ \citenamefont
  {Lo}}]{PhysRevA.72.012326}%
  \BibitemOpen
  \bibfield  {author} {\bibinfo {author} {\bibfnamefont {X.}~\bibnamefont
  {Ma}}, \bibinfo {author} {\bibfnamefont {B.}~\bibnamefont {Qi}}, \bibinfo
  {author} {\bibfnamefont {Y.}~\bibnamefont {Zhao}}, \ and\ \bibinfo {author}
  {\bibfnamefont {H.-K.}\ \bibnamefont {Lo}},\ }\href {\doibase
  10.1103/PhysRevA.72.012326} {\bibfield  {journal} {\bibinfo  {journal} {Phys.
  Rev. A}\ }\textbf {\bibinfo {volume} {72}},\ \bibinfo {pages} {012326}
  (\bibinfo {year} {2005})}\BibitemShut {NoStop}%
\bibitem [{\citenamefont {Bennett}\ \emph {et~al.}(1984)\citenamefont
  {Bennett}, \citenamefont {Brassard} \emph {et~al.}}]{bennett1984quantum}%
  \BibitemOpen
  \bibfield  {author} {\bibinfo {author} {\bibfnamefont {C.}~\bibnamefont
  {Bennett}}, \bibinfo {author} {\bibfnamefont {G.}~\bibnamefont {Brassard}},
  \emph {et~al.},\ }in\ \href@noop {} {\emph {\bibinfo {booktitle} {Proceedings
  of IEEE International Conference on Computers, Systems and Signal
  Processing}}},\ Vol.\ \bibinfo {volume} {175}\ (\bibinfo {organization}
  {Bangalore, India},\ \bibinfo {year} {1984})\BibitemShut {NoStop}%
\bibitem [{\citenamefont {Bennett}\ \emph {et~al.}(1992)\citenamefont
  {Bennett}, \citenamefont {Brassard},\ and\ \citenamefont
  {Mermin}}]{bennett_quantum_1992}%
  \BibitemOpen
  \bibfield  {author} {\bibinfo {author} {\bibfnamefont {C.~H.}\ \bibnamefont
  {Bennett}}, \bibinfo {author} {\bibfnamefont {G.}~\bibnamefont {Brassard}}, \
  and\ \bibinfo {author} {\bibfnamefont {N.~D.}\ \bibnamefont {Mermin}},\
  }\href {\doibase 10.1103/PhysRevLett.68.557} {\bibfield  {journal} {\bibinfo
  {journal} {Phys. Rev. Lett.}\ }\textbf {\bibinfo {volume} {68}},\ \bibinfo
  {pages} {557{\textendash}559} (\bibinfo {year} {1992})}\BibitemShut {NoStop}%
\bibitem [{\citenamefont {Lo}\ \emph {et~al.}(2005{\natexlab{b}})\citenamefont
  {Lo}, \citenamefont {Chau},\ and\ \citenamefont {Ardehali}}]{Lo2005}%
  \BibitemOpen
  \bibfield  {author} {\bibinfo {author} {\bibfnamefont {H.~K.}\ \bibnamefont
  {Lo}}, \bibinfo {author} {\bibfnamefont {H.}~\bibnamefont {Chau}}, \ and\
  \bibinfo {author} {\bibfnamefont {M.}~\bibnamefont {Ardehali}},\ }\href
  {\doibase 10.1007/s00145-004-0142-y} {\bibfield  {journal} {\bibinfo
  {journal} {Journal of Cryptology}\ }\textbf {\bibinfo {volume} {18}},\
  \bibinfo {pages} {133} (\bibinfo {year} {2005}{\natexlab{b}})}\BibitemShut
  {NoStop}%
\bibitem [{\citenamefont {Min\'a\ifmmode~\check{r}\else \v{r}\fi{}}\ \emph
  {et~al.}(2012)\citenamefont {Min\'a\ifmmode~\check{r}\else \v{r}\fi{}},
  \citenamefont {de~Riedmatten},\ and\ \citenamefont {Sangouard}}]{minar}%
  \BibitemOpen
  \bibfield  {author} {\bibinfo {author} {\bibfnamefont {J.~c.~v.}\
  \bibnamefont {Min\'a\ifmmode~\check{r}\else \v{r}\fi{}}}, \bibinfo {author}
  {\bibfnamefont {H.}~\bibnamefont {de~Riedmatten}}, \ and\ \bibinfo {author}
  {\bibfnamefont {N.}~\bibnamefont {Sangouard}},\ }\href {\doibase
  10.1103/PhysRevA.85.032313} {\bibfield  {journal} {\bibinfo  {journal} {Phys.
  Rev. A}\ }\textbf {\bibinfo {volume} {85}},\ \bibinfo {pages} {032313}
  (\bibinfo {year} {2012})}\BibitemShut {NoStop}%
\bibitem [{\citenamefont {Weinfurter}(1994)}]{0295-5075-25-8-001}%
  \BibitemOpen
  \bibfield  {author} {\bibinfo {author} {\bibfnamefont {H.}~\bibnamefont
  {Weinfurter}},\ }\href {http://stacks.iop.org/0295-5075/25/i=8/a=001}
  {\bibfield  {journal} {\bibinfo  {journal} {EPL (Europhysics Letters)}\
  }\textbf {\bibinfo {volume} {25}},\ \bibinfo {pages} {559} (\bibinfo {year}
  {1994})}\BibitemShut {NoStop}%
\bibitem [{\citenamefont {Calsamiglia}\ and\ \citenamefont
  {L{\"u}tkenhaus}(2001)}]{calsamiglia2001maximum}%
  \BibitemOpen
  \bibfield  {author} {\bibinfo {author} {\bibfnamefont {J.}~\bibnamefont
  {Calsamiglia}}\ and\ \bibinfo {author} {\bibfnamefont {N.}~\bibnamefont
  {L{\"u}tkenhaus}},\ }\href@noop {} {\bibfield  {journal} {\bibinfo  {journal}
  {Applied Physics B: Lasers and Optics}\ }\textbf {\bibinfo {volume} {72}},\
  \bibinfo {pages} {67} (\bibinfo {year} {2001})}\BibitemShut {NoStop}%
\bibitem [{\citenamefont {Kok}\ and\ \citenamefont
  {Lovett}(2010)}]{kok_introduction_2010}%
  \BibitemOpen
  \bibfield  {author} {\bibinfo {author} {\bibfnamefont {P.}~\bibnamefont
  {Kok}}\ and\ \bibinfo {author} {\bibfnamefont {B.~W.}\ \bibnamefont
  {Lovett}},\ }\href@noop {} {\emph {\bibinfo {title} {Introduction to optical
  quantum information processing}}}\ (\bibinfo  {publisher} {Cambridge
  University Press},\ \bibinfo {address} {Cambridge},\ \bibinfo {year}
  {2010})\BibitemShut {NoStop}%
\bibitem [{\citenamefont {Collins}\ \emph {et~al.}(2007)\citenamefont
  {Collins}, \citenamefont {Jenkins}, \citenamefont {Kuzmich},\ and\
  \citenamefont {Kennedy}}]{Collins2007}%
  \BibitemOpen
  \bibfield  {author} {\bibinfo {author} {\bibfnamefont {O.}~\bibnamefont
  {Collins}}, \bibinfo {author} {\bibfnamefont {S.}~\bibnamefont {Jenkins}},
  \bibinfo {author} {\bibfnamefont {A.}~\bibnamefont {Kuzmich}}, \ and\
  \bibinfo {author} {\bibfnamefont {T.}~\bibnamefont {Kennedy}},\ }\href
  {\doibase 10.1103/PhysRevLett.98.060502} {\bibfield  {journal} {\bibinfo
  {journal} {Physical Review Letters}\ }\textbf {\bibinfo {volume} {98}},\
  \bibinfo {pages} {060502} (\bibinfo {year} {2007})}\BibitemShut {NoStop}%
\bibitem [{\citenamefont {Bernardes}\ \emph {et~al.}(2011)\citenamefont
  {Bernardes}, \citenamefont {Praxmeyer},\ and\ \citenamefont {van
  Loock}}]{NadjaHybrid}%
  \BibitemOpen
  \bibfield  {author} {\bibinfo {author} {\bibfnamefont {N.~K.}\ \bibnamefont
  {Bernardes}}, \bibinfo {author} {\bibfnamefont {L.}~\bibnamefont
  {Praxmeyer}}, \ and\ \bibinfo {author} {\bibfnamefont {P.}~\bibnamefont {van
  Loock}},\ }\href {\doibase 10.1103/PhysRevA.83.012323} {\bibfield  {journal}
  {\bibinfo  {journal} {Phys. Rev. A}\ }\textbf {\bibinfo {volume} {83}},\
  \bibinfo {pages} {012323} (\bibinfo {year} {2011})}\BibitemShut {NoStop}%
\bibitem [{\citenamefont {Abramowitz}\ \emph {et~al.}(1964)\citenamefont
  {Abramowitz} \emph {et~al.}}]{abramowitz1964handbook}%
  \BibitemOpen
  \bibfield  {author} {\bibinfo {author} {\bibfnamefont {M.~E.}\ \bibnamefont
  {Abramowitz}} \emph {et~al.},\ }\href@noop {} {\emph {\bibinfo {title}
  {Handbook of mathematical functions: with formulas, graphs, and mathematical
  tables}}},\ Vol.~\bibinfo {volume} {55}\ (\bibinfo  {publisher} {Courier
  Dover Publications},\ \bibinfo {year} {1964})\BibitemShut {NoStop}%
\bibitem [{\citenamefont {Abruzzo}\ \emph {et~al.}(2013)\citenamefont
  {Abruzzo}, \citenamefont {Bratzik}, \citenamefont {Bernardes}, \citenamefont
  {Kampermann}, \citenamefont {van Loock},\ and\ \citenamefont
  {Bru\ss{}}}]{abruzzo2012quantum}%
  \BibitemOpen
  \bibfield  {author} {\bibinfo {author} {\bibfnamefont {S.}~\bibnamefont
  {Abruzzo}}, \bibinfo {author} {\bibfnamefont {S.}~\bibnamefont {Bratzik}},
  \bibinfo {author} {\bibfnamefont {N.~K.}\ \bibnamefont {Bernardes}}, \bibinfo
  {author} {\bibfnamefont {H.}~\bibnamefont {Kampermann}}, \bibinfo {author}
  {\bibfnamefont {P.}~\bibnamefont {van Loock}}, \ and\ \bibinfo {author}
  {\bibfnamefont {D.}~\bibnamefont {Bru\ss{}}},\ }\href {\doibase
  10.1103/PhysRevA.87.052315} {\bibfield  {journal} {\bibinfo  {journal} {Phys.
  Rev. A}\ }\textbf {\bibinfo {volume} {87}},\ \bibinfo {pages} {052315}
  (\bibinfo {year} {2013})}\BibitemShut {NoStop}%
\bibitem [{\citenamefont {Jofre}\ \emph {et~al.}(2010)\citenamefont {Jofre},
  \citenamefont {Gardelein}, \citenamefont {Anzolin}, \citenamefont
  {Molina-Terriza}, \citenamefont {Torres}, \citenamefont {Mitchell},\ and\
  \citenamefont {Pruneri}}]{5504020}%
  \BibitemOpen
  \bibfield  {author} {\bibinfo {author} {\bibfnamefont {M.}~\bibnamefont
  {Jofre}}, \bibinfo {author} {\bibfnamefont {A.}~\bibnamefont {Gardelein}},
  \bibinfo {author} {\bibfnamefont {G.}~\bibnamefont {Anzolin}}, \bibinfo
  {author} {\bibfnamefont {G.}~\bibnamefont {Molina-Terriza}}, \bibinfo
  {author} {\bibfnamefont {J.~P.}\ \bibnamefont {Torres}}, \bibinfo {author}
  {\bibfnamefont {M.~W.}\ \bibnamefont {Mitchell}}, \ and\ \bibinfo {author}
  {\bibfnamefont {V.}~\bibnamefont {Pruneri}},\ }\href {\doibase
  10.1109/JLT.2010.2056673} {\bibfield  {journal} {\bibinfo  {journal}
  {Lightwave Technology, Journal of}\ }\textbf {\bibinfo {volume} {28}},\
  \bibinfo {pages} {2572 } (\bibinfo {year} {2010})}\BibitemShut {NoStop}%
\bibitem [{\citenamefont {Brassard}\ \emph {et~al.}(2000)\citenamefont
  {Brassard}, \citenamefont {L\"utkenhaus}, \citenamefont {Mor},\ and\
  \citenamefont {Sanders}}]{PhysRevLett.85.1330}%
  \BibitemOpen
  \bibfield  {author} {\bibinfo {author} {\bibfnamefont {G.}~\bibnamefont
  {Brassard}}, \bibinfo {author} {\bibfnamefont {N.}~\bibnamefont
  {L\"utkenhaus}}, \bibinfo {author} {\bibfnamefont {T.}~\bibnamefont {Mor}}, \
  and\ \bibinfo {author} {\bibfnamefont {B.~C.}\ \bibnamefont {Sanders}},\
  }\href {\doibase 10.1103/PhysRevLett.85.1330} {\bibfield  {journal} {\bibinfo
   {journal} {Phys. Rev. Lett.}\ }\textbf {\bibinfo {volume} {85}},\ \bibinfo
  {pages} {1330} (\bibinfo {year} {2000})}\BibitemShut {NoStop}%
\end{thebibliography}%

\end{document}